\documentclass[aps,prd,floats,nofootinbib,preprintnumbers,11pt]{revtex4}
\usepackage{epsfig}
\usepackage{amsmath}
\usepackage{amssymb}
\usepackage{amsthm}
\usepackage{longtable}
\usepackage{array}
\usepackage[T1]{fontenc}
\usepackage[latin9]{inputenc}
\usepackage{color}
\usepackage{graphicx}
\usepackage{esint}

\begin{document}

\preprint{NUHEP-TH/08-04}

\title{Minimally Allowed $\beta\beta0\nu$ Rates Within an Anarchical Framework}

\author{James Jenkins}
\affiliation{Northwestern University, Department of Physics \&
Astronomy, 2145 Sheridan Road, Evanston, IL~60208, USA}

\email{jjenkins6@lanl.gov}

\begin{abstract}
Neutrinoless double beta decay $(\beta\beta0\nu)$ is the only realistic
probe of the Majorana nature of the neutrino. In the standard picture,
its rate is proportional to $m_{ee}$, the $e-e$ element of the Majorana
neutrino mass matrix in the flavor basis. I explore minimally allowed
$m_{ee}$ values within the framework of mass matrix anarchy where
neutrino parameters are defined statistically at low energies. Distributions
of mixing angles are well defined by the Haar integration measure,
but masses are dependent on arbitrary weighting functions and boundary
conditions. I survey the integration measure parameter space and find
that for sufficiently convergent weightings, $m_{ee}$ is constrained
between $(0.01-0.4)$ eV at 90\% confidence. Constraints from neutrino
mixing data lower these bounds. Singular integration measures allow
for arbitrarily small $m_{ee}$ values with the remaining elements
ill-defined, but this condition constrains the flavor structure of
the model's ultraviolet completion. $\beta\beta0\nu$ bounds below
$m_{ee}\sim5\times10^{-3}$ eV should indicate symmetry in the lepton
sector, new light degrees of freedom or the Dirac nature of the neutrino.
\end{abstract}
\maketitle

\section{Introduction \label{sec:Motivation}}

The observation of nonzero neutrino mass and mixing via flavor oscillations
is the first terrestrial evidence of physics beyond the Standard Model
(SM) of particle physics. See \citep{NuRev:AndreSoWhat,NuRev:AndreTASI,NuRev:TheoryWhitePaper}
for a review of neutrino physics. This discovery provides useful insight
into the nature of new high scale phenomena and introduces many important
questions. Chief among these is the charge conjugation properties
(Dirac vs. Majorana) of the neutrino. First, one may enlarge the SM
field content to include one or more gauge singlets $N$ (right handed
neutrinos) and give the neutrino a Dirac mass similar to the other
fermions via coupling to the neutral SM Higgs boson: $\nu NH^{0}$.
On the other hand, possessing no unbroken gauge quantum numbers, the
neutrino may have a Majorana mass term which couples the neutrino
to its charge conjugate $\nu^{c}$, and thus renders the neutrino
equal to its own antiparticle. This breaks all nontrivial global symmetries.
In particular, a Majorana neutrino mass violates lepton number by
two units. The Dirac versus Majorana nature of the neutrino is an
important issue that must be addressed experimentally.

Neutrinoless double beta decay ($\beta\beta0\nu$) is an as yet unobserved
Lepton Number Violating (LNV) process that would unambiguously identify
the Majorana nature of the neutrino \citep{BlackBox}. In fact, except
in rare circumstances \citep{MyLNV}, $\beta\beta0\nu$ is the only
realistic hope of probing LNV in the near future \citep{Atre:UpperBoundsLNVProcesses,Lim:VarietyLNVProcessesRelated2MajNuMass}.
Given the importance of this process, it is useful to explore its
theoretical expectations over a broad range of scenarios. In the standard
picture, with no new light LNV degrees of freedom below the TeV scale,
$\beta\beta0\nu$ proceeds primarily via Majorana neutrino exchange.
In this case, its amplitude is proportional to $m_{ee}$, the $e-e$
element of the neutrino mass matrix in the flavor basis where the
charged leptons are diagonal.  While this relationship is generally
spoiled by new physics \citep{MyLNV}, it should still effectively
hold for sufficiently small $m_{ee}$ values.  While it is true that the smallest $\beta\beta0\nu$ rates may be dominated by high dimensional non-renormalizable operators there is an important feedback mechanism into $m_{ee}$. This follows from the
extended black box theorem \citep{ExtendedBlackBox} where a one to
one relationship is derived between $m_{ee}$ and the effective $m_{ee}^{eff}$
that governs $\beta\beta0\nu$ in the vanishing limit, such that $m_{ee}=0\leftrightarrow m_{ee}^{eff}=0$.
Thus, it is reasonable to assume that, when searching for small $\beta\beta0\nu$
rates, it is enough to study the behavior of $m_{ee}$.  This is only loose motivation for the present analysis as it is impossible to extract the exact rate below which this reasoning holds without assuming properties of the neutrino mass's ultraviolet completion.  In what follows I will assume the dominance of the light Majorana neutrino exchange mechanism parameterized by $m_{ee}$.  This is the most popular case.  Additionally, since other new physics has yet to be discovered, it is the minimal mechanism currently implied by direct observation.  Current experimental limits constrain the $\beta\beta0\nu$
half-life below $\sim10^{25}$ years, corresponding to $m_{ee}<0.35$
eV at 90\% confidence%
\footnote{The translation between measured half-life and $m_{ee}$ is not straight
forward, as it depends critically on isotope dependent nuclear matrix
element calculations where uncertainties currently range an order
of magnitude \citep{BB0nMatrixElUnc,Rodin:AssessmentUncertaintiesQRPABB0nNuclearMatrixElements,Menendez:DisassemblingNuclearMatrixElementsBB0n}.
This is likely to improve within the next several years. %
} \citep{IGEXBB0nResults,LatestResultsHeidelberMoscowBB0nExp,AbsoluteValuesNuMassStatusProspects,ImplicationsOfNuDataCirca2005}.
Next generation experiments are poised to extend this reach by roughly
an order of magnitude to $m_{ee}<0.05$ eV \citep{StrategiesForNextGenBB0nExperiments,AbsoluteValuesNuMassStatusProspects,Zuber:BB0nExperiments}. 

While it is true that $\beta\beta0\nu$ rates are below current sensitivities%
\footnote{A positive signal was reported by a subset of the HEIDELBERG-MOSCOW
group with a half-life near $1.19\times10^{25}$ years at $4.2\sigma$
confidence \citep{BB0nPosResultKlapdor}. I neglect this observation
in what follows awaiting conformation, except to point out that their
extracted $m_{ee}$ is well accommodated by the anarchy model of neutrino
mass.%
}, it is possible that they will be discovered by the next round of
experiments. In terms of the measured oscillation parameters, we are
only beginning to explore the interesting range dominated by the atmospheric
mass squared difference within the inverted and quasi-degenerate spectral
hierarchies \citep{Elliott:DoubleBetaDecay}. If $\beta\beta0\nu$
observation is ``right around the corner,'' in which case $m_{ee}$
is relatively large, it is unlikely that its rate is suppressed by
an approximate flavor symmetry. However, if bounds are pushed significantly
lower, it is quite reasonable the a small $m_{ee}$ is protected by
an appropriate symmetry mechanism \cite{MyMinBB0nFlavSym}. It is natural to wonder how small
it can be without the introduction of imposed mass matrix structure.
To this end, it is instructive to take the ``no structure'' limit
and consider $m_{ee}$ bounds, assuming the anarchy hypothesis \citep{Haba:2000be,Hall:1999sn}.  Similar reasoning was applied much earlier in the history of neutrino physics to probe the potential for larges mixing angles \cite{GoldmanHowLargeNuMixAngles}.
In this scenario, the underlying neutrino mass model is sufficiently
complicated, such that the mass matrix appears random at low energies
in any basis. In other words, there is effectively no difference between
the three light neutrino states. This leads to a distribution of observables
that must be treated statistically. In \citep{deGouvea:2003xe,Haba:2000be},
it was shown that the large mixing angles and small hierarchies of
the neutrino sector are consistent with anarchy, provided $\theta_{13}$
is not too small, whereas the CKM matrix is inconsistent with anarchy,
as expected by pure inspection of its structure%
\footnote{These claims were questioned and explored by analysis discussed in
\citep{Espinosa:2003qz} and \citep{Hirsch:2001mw} but the overall
consistency between anarchy and current neutrino data still holds.%
}. In this analysis, the marginalized mixing angle distribution functions
are well defined in terms of the Haar measure invariant under the
$U(3)$ group. It is not as straightforward to consider questions
involving mass eigenvalues since one may include arbitrary $U(3)$
invariant weighting functions into the integration measure \citep{Haba:2000be}
and boundary conditions. Here, I survey this added layer of ambiguity
and derive the smallest allowed mass matrix element consistent with
anarchy.

In what follows, I analyze expectations for $m_{ee}$ within the anarchy
picture of neutrino mass generation. The goal is to determine how
small/large one mass matrix element may be from the others within
an anarchical framework. In section \ref{sec:Mass-Matrix-Anarchy},
I introduce the formalism and notation employed throughout the analysis.
Using the Kolmogorov-Smirnov (KS) goodness of fit test, I scan the
parameter space of measures defined by both simple polynomial and
divergent $U(3)$ invariant functions with ``spherical'' boundary
conditions in Subsection \ref{sub:PolyMeasure}. Here, I also comment
on modifications induced by the use of nontrivial boundary conditions.
In Subsection \ref{sub:BB0N-Rates-And}, I connect these results to
the case of realistic neutrino mixing. I conclude in Section \ref{sec:Conclusion}
with a summary of my results and comment on the impact of future experimental
data.

\section{Mass Matrix Anarchy and $\beta\beta0\nu$ \label{sec:Mass-Matrix-Anarchy}}

I parameterize the complex, symmetric, three neutrino Majorana mass
matrix as\begin{equation}
m_{\alpha\beta}=ma_{\alpha\beta}=mr_{\alpha\beta}e^{i\phi_{\alpha\beta}},\label{eq:malphabeta}\end{equation}
where the dimensionless complex parameters $a_{\alpha\beta}\equiv(a)_{\alpha\beta}$
define the structure of the matrix and are constructed to have a magnitude
$r_{\alpha\beta}$ and phase $\phi_{\alpha\beta}$. The latter contains
the familiar Majorana and Dirac phases of the neutrino mixing matrix
in various linear combinations. Three of these six phases may be rotated
away as unphysical with appropriate transformations, but are included in this analysis
without loss of generality.  See for example \citep{deGouvea:2008nm,Jenkins:RephasingInvariantsQuarkLeptonMixingMatrices} and references therein. A dimensionful factor of $m$ is pulled
out to carry the scale of neutrino masses. There is ambiguity in the
factorization of $m$ and $r_{\alpha\beta}$. To be concrete, I define
$m$ such that the resulting $r_{\alpha\beta}$ matrix elements have
maximal magnitudes defined by boundary conditions subject to anarchy
constraints. The average $r_{\alpha\beta}$ values should be near
unity. In other words, $r_{\alpha\beta}$ is a generally $\mathcal{O}(1)$
matrix up to some deviations described by the anarchy hypothesis.
The overall neutrino mass scale $m$ is inferred from experiment and
included into the analysis by hand. Currently, $m$ is bounded at
$0.05$ eV from below by the atmospheric mass squared difference \citep{ImplicationsOfNuDataCirca2005,StatusGlobFitsMaltoni07}
and from above at roughly $1$ eV by cosmological data \citep{ImplicationsOfNuDataCirca2005}.
In the remainder of this paper I refer to these as the hierarchical
and quasi-degenerate limits, respectively. These are realized when
the lightest mass eigenvalue squared is less than (hierarchical) or
greater than (quasi-degenerate) the mass squared differences. In what
follows, within the anarchy picture, one may only extract limits on
the deviations of $r_{\alpha\beta}$ from its average value, or equivalently,
bound the ratio of mass matrix elements. 

Imposing the anarchy hypothesis implies that the matrix elements of
Eq.~\eqref{eq:malphabeta} are distributed randomly in any basis or,
more precisely, the probability distribution of each $a_{\alpha\beta}$
is invariant under arbitrary unitary rotations. Starting from the
diagonal mass basis, the flavor basis (as well as any other physical
basis) is found by a random rotation, and one may calculate the probability
distribution of any matrix structure. Notice that the obvious distinction
of the diagonal mass basis renders it a very improbable structure
which may be understood in terms of a sampled ensemble of matrices.
The small chance of landing on it via random rotation does not preclude
its existence.

Assuming three light neutrinos, the condition of $U(3)$ invariance
imposes strict conditions on the total distribution function $G(a_{\alpha\beta})$
universally associated with each element. This quantity may only be
a function of\[
\det(a)=\epsilon_{ijk}r_{1i}r_{2j}r_{3k}e^{i\left(\phi_{1i}+\phi_{2j}+\phi_{3k}\right)}\]
and the purely radial quantity\[
{\rm tr}(a^{\dagger}a)=r_{11}^{2}+r_{22}^{2}+r_{33}^{2}+2r_{12}^{2}+2r_{13}^{2}+2r_{23}^{2}.\]
Working in these polar coordinates, the marginalized probability distribution
of any single element's magnitude, say $|a_{11}|=r_{11}\equiv r$,
is obtained by integrating over the other radial coordinates and all
phases, subject to some (also invariant) integration limits $l=l[{\rm tr}(a^{\dagger}a),\det(a)]$.
Thus, the $r$-coordinate distribution may be written in terms of
the arbitrary functionals $G$ and $l$, as 

\begin{equation}
g(r)=\int_{0}^{l\left[{\rm tr}(a^{\dagger}a),\det(a)\right]}G\left[{\rm tr}(a^{\dagger}a),\det(a)\right]\prod_{\alpha\leq\beta,\beta\neq1}r_{\alpha\beta}dr_{\alpha\beta}\prod_{\alpha\leq\beta}d\phi_{\alpha\beta}.\label{eq:gofr}\end{equation}
At this point, $G$ may be thought of as a weighting function in the
integration measure. As a probability, $g(r)$ is positive and normalized
as $\int_{0}^{1}g(r)rdr=1$. This implies that matrix elements do
not have independent distribution functions, so that the probability
of structures defined by magnitudes $r_{\alpha\beta}$ cannot be expressed
as a product of $g(r_{\alpha\beta})$. Rather, one must calculate
the generalized version Eq.~\eqref{eq:gofr}. The cumulative probability
distribution that the radial magnitude is less than some $r$ is given
by

\begin{equation}
F(r)=\int_{0}^{r}g(r^{\prime})r^{\prime}dr^{\prime}.\label{eq:Fofr}\end{equation}
With this, the $C$ confidence interval limits are found from 

\begin{equation}
{\rm max}\left[F(r),1-F(r)\right]-\frac{C+1}{2}=0,\label{eq:ConfidenceLimits}\end{equation}
which yields two solutions interpreted as the extreme $r$ values
allowed at C confidence. One may trivially generalize this procedure
to $n$ neutrino states with no qualitative changes. It reflects the
mechanism behind the KS goodness of fit test and yields the same results
for one free variable.

\subsection{Simple Polynomial Measures\label{sub:PolyMeasure}}

Consider a one-dimensional marginalized probability distribution function,
limited by $l={\rm Tr}(a^{\dagger}a)\leq3^{2}$. This choice is not
unique but does make physical sense, as it implies that each matrix
element lives within a three-dimensional sphere. This is clear upon
rotation to the diagonal mass eigenbasis. I chose a non-unit radius
for convenience to reflect the physical definition of the mass $m$,
but this may be changed, provided a corresponding scaling of $r$.
With this choice, the universal $\mathcal{O}(1)$ matrix structures
indicative of anarchy saturates the upper integration bound. This
is easy to understand, as sample volumes at large radii dominates
that at smaller radii. Thus, $r_{\alpha\beta}\in[0,3]$, with the
most probable values naively expected near one. It is reasonable to
assume the weighting functional may be expanded in a double Taylor
series as $G\left[{\rm tr}(a^{\dagger}a),\det(a)\right]=\sum_{p,q}^{\infty}c_{pq}{\rm tr}(a^{\dagger}a)^{p}{\rm det}(a)^{q}$.
Upon integrating over the complex phases it is clear that all nonzero
powers of ${\rm det}(a)$ vanish by symmetry. However, this is only
true with $\phi_{\alpha\beta}$ independent integration limits. A
nontrivial dependence on $\det(a)$ is possible within this framework
but requires a departure from the physically motivated spherical boundary
conditions. An example case of polynomial dependence reveals that
such limits have little impact on the marginalized distribution function
for small values of $r$, which is the primary concern of this analysis.

Thus, it is enough to consider only linear combinations of\begin{equation}
G_{p}\left[{\rm tr}(a^{\dagger}a),\det(a)\right]={\rm Tr}(a^{\dagger}a)^{p}=\left(r^{2}+r_{22}^{2}+r_{33}^{2}+2r_{12}^{2}+2r_{23}^{2}+2r_{13}^{2}\right)^{p}.\label{eq:MeasureP}\end{equation}
In this case, it is simple to obtain the cumulative distribution function 

\begin{equation}
F_{p}(\tilde{r})=(6+p)\tilde{r}^{2}\left[\frac{5}{5+p}-\frac{10\tilde{r}^{2}}{4+p}+\frac{10\tilde{r}^{4}}{3+p}-\frac{5\tilde{r}^{6}}{2+p}+\frac{\tilde{r}^{8}}{1+p}-\frac{120\tilde{r}^{10+2p}}{(6+p)(5+p)(4+p)(3+p)(2+p)(1+p)}\right]\label{eq:Fpofr}\end{equation}
via the procedure outlined in Eq.~\eqref{eq:gofr} and Eq.~\eqref{eq:Fofr}
in terms of the scaled magnitude $\tilde{r}\equiv r/3$ that lives
in the unit interval. This is defined for all $p>-6$. It is easy
to see that all other pole divergences cancel pairwise out of this
expression. 

\begin{figure}
\includegraphics[scale=0.75]{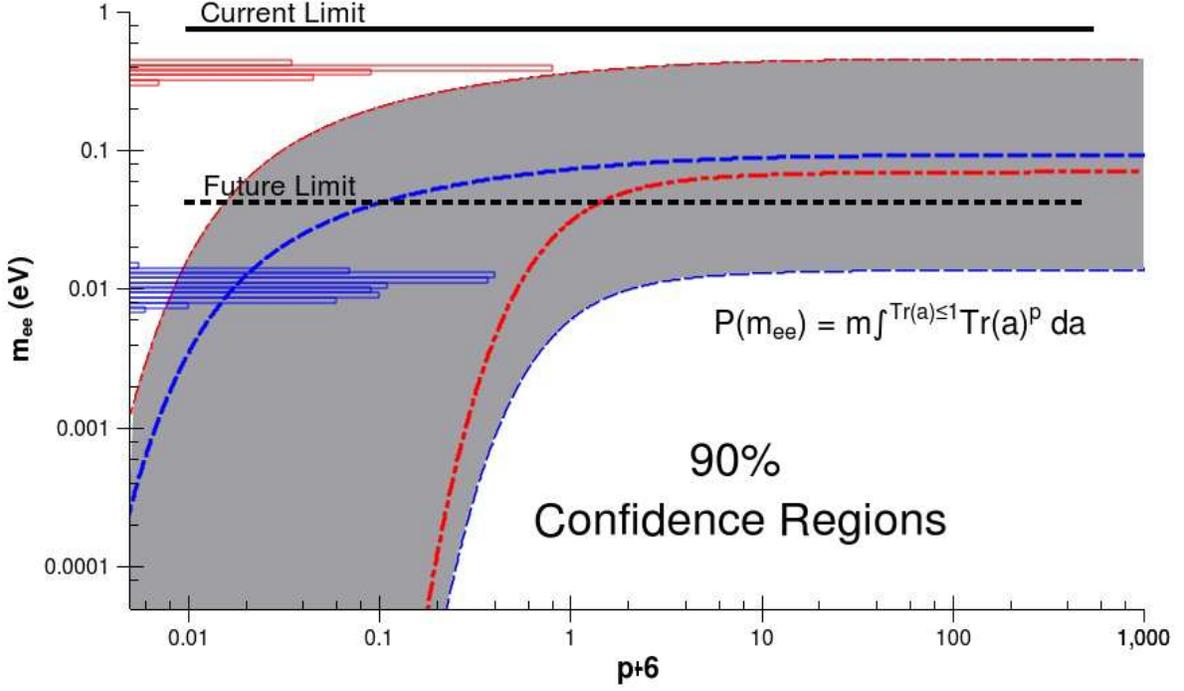}

\caption{Anarchy allowed $m_{ee}$ limits at 90\% confidence as a function
of $p+6$ defined by Eq.~\eqref{eq:MeasureP} (see text for details).
Shaded region is total confidence interval while the blue (dash) and
red (dash-dot) bounds indicate the hierarchical and quasi-degenerate
neutrino mass spectra, respectively. $m_{ee}\rightarrow0$ as $p\rightarrow-6$.
Red and blue horizontal histograms show allowed upper and lower $m_{ee}$
bounds for a measure series expansion sample space for quasi-degenerate
and hierarchical spectra, respectively. Current and future $\beta\beta0\nu$
bounds are also shown.}
\label{fig:Contour90}
\end{figure}

Combining Eq.~\eqref{eq:Fpofr} and Eq.~\eqref{eq:ConfidenceLimits}
yields limits on the ratio of matrix elements $r$. Upon multiplication
by the current experimentally allowed neutrino mass scale range $0.05{\rm eV}<m<1{\rm eV}$,
one may obtain the two sided 90\% confidence limits of $m_{ee}$ (or
any other matrix element) as a function of $p$ within the anarchy
framework. This is shown in the gray contour of Figure \ref{fig:Contour90}
along with the current and future $\beta\beta0\nu$ reach denoted
by horizontal dotted and dashed lines, respectively. For convenience,
the 90\% allowed region is decomposed into the two extreme cases of
hierarchical and quasi-degenerate mass spectra bounded by the blue
dashed and red dotted-dashed curves, respectively. Given a particular
neutrino mass scale, the allowed $m_{ee}$ range spans scarcely an
order of magnitude. Much of the breadth of the totally allowed grey
region in Figure \ref{fig:Contour90} comes from uncertainty in $m$.
Therefore, an absolute neutrino mass measurement should significantly
tighten the anarchy prediction of $m_{ee}$.

For $p>-5$ the curves of Figure \ref{fig:Contour90} are relatively
level yielding $m_{ee}$ in the range (0.01-0.1) eV and (0.05-0.5)
eV for the hierarchical and quasi-degenerate spectra. This is easy
to understand, as $\tilde{r}\leq1$ and Eq.~\eqref{eq:Fpofr} may be
truncated at quadratic order%
\footnote{This is only true for $p>-5$ as, in this limit, the quadratic terms
of Eq.~\eqref{eq:Fpofr} diverge and thus cancel. In this case one
must truncate at quartic order.%
}. Hence, $F_{p}(r)\rightarrow\frac{5}{9}\frac{6+p}{5+p}r^{2}$ yielding
a $C$ confidence limit for $r$ between $3\sqrt{(1\pm C)\frac{5+p}{10(6+p)}}$.
This approximation works very well for lower bounds but receives up
to 5\% corrections for the upper 90\% confidence limits. At this point,
it is instructive to consider the generalized $n$ neutrino scenario
for analogous weightings and boundary conditions $l={\rm Tr}(a^{\dagger}a)\leq n^{2}$.
In the small $\tilde{r}=r/n$ approximation, the cumulative distribution
function is given by $F_{p}^{(n)}\approx r^{2}\frac{(n-1)(n+2)}{2n^{2}}\left\{ \frac{2p+n^{2}+n}{2p+n^{2}+n-2}\right\} $,
which yields confidence limits between $n\sqrt{(1\pm C)\frac{2p+n^{2}+n-2}{(n-1)(n+2)(2p+n^{2}+n)}}$.
At large $n$, the matrix elements become independent with distributions
bounded by $\sqrt{1\pm}C$. 

I now consider general well behaved measure functions as series expansions
of ${\rm Tr}(a^{\dagger}a)^{p}$ weighted by arbitrary coefficients
$c_{p}$, which by a trivial extension of Eq.~\eqref{eq:Fpofr} results
in \begin{equation}
F(\tilde{r})=\frac{1}{\sum_{m=0}^{\infty}\frac{c_{m}}{m+6}}\sum_{p=0}^{\infty}\frac{c_{p}}{6+p}F_{p}(\tilde{r}).\label{eq:FSeriesofr}\end{equation}
Here, one coefficient is always redundant, as it may be factored out
of both the numerator and denominator and ultimately canceled. Due
to this normalization condition, only those terms with comparable
weightings having large destructive/constructive interference contribute
to deviations from the limit contours of Figure \ref{fig:Contour90}.
This can be seen in the small $\tilde{r}$ approximation $F(\tilde{r})\approx5\tilde{r}^{2}\sum_{p=0}^{\infty}\frac{c_{p}}{p+5}/\sum_{m=0}^{\infty}\frac{c_{m}}{m+6}$.
Uniformly scanning the parameters of Eq.~\eqref{eq:FSeriesofr}, truncated
at $p=4$ within the range $c_{p}\in[-50,50]$, I sample the $m_{ee}$
bounds obtained from $10^{10}$ sample functions. Results are shown
in the blue and red horizontal histograms of Figure \ref{fig:Contour90}
for the hierarchical and quasi-degenerate mass scales, respectively.
These arbitrarily normalized distributions are highly dependent on
the sampling procedure. Consequently, the upper and lower endpoints
are the only useful quantities. Due to the small $\tilde{r}$ approximate
behavior, the distribution of upper bounds is much wider than that
of the lower.  The behavior of these histograms are in analogy with the work of \cite{Haba:2000be} where random mass matrices were generated and studied using a linear measure function and cubic boundary conditions.  In that case, mass eigenvalues were histogrammed as opposed to the combined quantity $m_{ee}$.  

The behavior of $F_{p}(\tilde{r})$ for $p\rightarrow-6$ is easy
to understand, but surprising from the mass matrix anarchy viewpoint.
Approaching this lower limit, one finds that $\lim_{p\rightarrow-6}F_{p}(\tilde{r})\approx\tilde{r}^{2(p+6)}\approx1+(p+6)\ln\tilde{r}^{2}$,
implying that the marginalized probability distribution $g(r)\rightarrow\delta(r).$
This shows that at least one mass matrix element may be arbitrarily
small, provided a sufficiently divergent integration measure. Furthermore,
this must hold true in any basis obtained by a random rotation from
the diagonal mass basis. With this, one finds the C confidence region
bounded between $e^{-\frac{1\pm C}{4(6+p)}}$, which goes rapidly
to zero, as seen in Figure \ref{fig:Contour90}. It remains to check
the behavior of the other, marginalized matrix elements in this limit.
Given two magnitudes, $r$ and $s$, one may calculate the marginalized
probability distribution function $g_{p}(r,s)$ and the cumulative
distribution function $F_{p}(r,s)$ parameterized by $p$ in the integration
measure of Eq.~\eqref{eq:MeasureP}. In the limit $p\rightarrow-6$,
when $\tilde{r}$ is small, per the above argument, \[
F_{p}(r,s)\approx\tilde{r}^{2(p+6)}-(p+6)\frac{r^{2}}{s^{2}}=F_{p}(r)-(p+6)\frac{r^{2}}{s^{2}}.\]
When $r$ is within its C confidence limits, defined by $F_{p}(r)\in(1\pm C)/2$,
then $F_{p}(r,s)\in(1\pm C)/2$ is also satisfied up to small perturbations.
The coordinate $s$ is virtually unconstrained. Thus, given $p\rightarrow-6$,
at least one texture zero is guaranteed and all other matrix elements
vary freely and are (almost) independent of the integration measure.
Similar statements apply to other sufficiently divergent poles in
the measure. For example, given $G[{\rm Tr}(a^{\dagger}a)]=(B^{2}-{\rm Tr}(a^{\dagger}a))^{p}$
in the appropriate $p$ limit, one radial magnitude goes to $B$ while
the rest are unconstrained by the theory.

\subsection{Mass Matrix Anarchy and Neutrino Mixing \label{sub:BB0N-Rates-And}}

Thus far, I have explored the relation: Neutrino mixing data is consistent
with the anarchy hypothesis... Given neutrino anarchy, what are the
allowed mass matrix structures in an arbitrary basis? The allowed
$m_{ee}$ range is then easily extracted. This is the proper treatment,
but it does not address if/when the derived structures accommodate
the neutrino data. Alternately, one might consider the anarchical
$m_{ee}$ distribution constrained by current oscillation phenomenology.
This will not yield the same range as the previous analysis. As noted
in \citep{deGouvea:2003xe}, regarding the preferred value of the
reactor mixing angle, given the consistency of neutrino data and anarchy,
one would expect an expanded confidence region for unknown quantities
such as the CP phases and $\theta_{13}$. In other words, a large
amount of (accidental) structure can appear among these parameters
while still maintaining consistency with anarchy. In the standard
parametrization, the composite mass matrix element $m_{ee}$ magnitude
is given by\begin{equation}
|m_{ee}|=m\left|r_{1}e^{i\phi_{1}}\cos^{2}\theta_{12}\cos^{2}\theta_{13}+r_{2}e^{i\phi_{2}}\sin^{2}\theta_{12}\cos^{2}\theta_{13}+r_{3}e^{i(\phi_{3}-\delta)}\sin^{2}\theta_{13}\right|,\label{eq:Meediag}\end{equation}
where an overall mass scale $m$ is factored out for consistency with
the preceding analysis. Here $r_{i}$ and $\phi_{i}$ make up the
complex eigenvalues of the flavor basis mass matrix Eq.~\eqref{eq:malphabeta},
diagonalized by the mixing angles $\theta_{12}$, $\theta_{23}$,
$\theta_{13}$ and $\delta$ in the usual way. In this quantity, a
single Majorana phase may be removed as unphysical so the resulting
expression depends on only two phase linear combinations. The mixing
parameters are distributed according to the Haar measure and the radial
magnitudes according to the complex Majorana mass eigenvalue measure
given in \citep{Haba:2000be}, weighted by an invariant function analogous
to Eq.~\eqref{eq:MeasureP}. $m_{ee}$ is the summed convolution of
these distributions. It is well known that, given current data, Eq.~\eqref{eq:Meediag}
can vanish, provided the normal neutrino mass ordering and particular
relationships among the phases and mass scale \citep{BB0nFutureNuOscPrecisionExp,MyNonOscHier,ImplicationsOfNuDataCirca2005}.
One would then expect the allowed $m_{ee}$ range to extend lower
in the normal than the inverted hierarchy. 

\begin{table}
\begin{tabular}{|c|c|c|c|}
\hline 
Name & Parameter Combination & Value & $1\sigma$ Uncertainty\tabularnewline
\hline
\hline 
$\Delta m^2_S$ & $m_2^2 - m_1^2$ & $7.65\times 10^{-5}~{\rm eV^2}$ & $0.22\times 10^{-5}~{\rm eV^2}$\tabularnewline
\hline 
 $\Delta m^2_A$ & $|m_3^2-m_2^2|$ & $2.40\times 10^{-3}~{\rm eV^2}$ & $0.12\times 10^{-3}~{\rm eV^2}$\tabularnewline
\hline 
 $\sin\theta_S$ & $\sin\theta_{12}$ & $0.551$ & $0.017$ \tabularnewline
\hline 
 $\sin\theta_A$ & $\sin\theta_{23}$ & $0.707$ & $0.046$\tabularnewline
\hline 
 $\sin\theta_R$ & $\sin\theta_{13}$ & $0.1$ & $<0.14$ \tabularnewline
\hline 

\end{tabular}
\caption{Summary table of current neutrino results together with naming conventions and parameter definitions.  Columns three and four list best fit central parameter values and $1\sigma$ uncertainties.  These were adapted from the global oscillation analysis of \cite{ThreeFlavorOscUpdate}.  The central values are used as input into the analysis of subsection \ref{sub:BB0N-Rates-And}.}
\label{tab:Constraints}
\end{table}

I point out that the preceding analysis could have been done from
this perspective, with separated mass and mixing parameters, but was
easier done in the flavor basis. Within a realistic neutrino mixing
framework constrained by the data shown in Table \ref{tab:Constraints}, the use of the convoluted Eq.~\eqref{eq:Meediag} is convenient
since we already know many best fit bounds from oscillation searches
\citep{StatusGlobFitsMaltoni07,ImplicationsOfNuDataCirca2005}. Marginalizing
the convoluted $m_{ee}$ distribution over the unknown phases as well
as the neutrino mixing angle uncertainties with a polynomial weighting
function, I obtain the normalized $m_{ee}$ cumulative distribution
function.  Thus, this quantity is the normalized term by term convolution of angular distribution functions weighted by the Haar measure and the polynomial measured mass eigenstate distributions.  The experimentally allowed mixing angle ranges may be substituted into this function. For simplicity, I consider the normal and inverted neutrino
mass spectra separately in terms of only the lowest free mass eigenvalue
with the others related to it by the measured mass squared differences.
I find that for all non-singular weightings, $m_{ee}>4.4\times10^{-4}$
eV and $m_{ee}>1.6\times10^{-2}$ eV at 90\% confidence for the normal
and inverted hierarchies, respectively. For the inverted hierarchy,
this bound coincides roughly with the smallest possible inverted $m_{ee}$
value obtained from Eq.~\eqref{eq:Meediag} evaluated at $r_{3}=0$.
As expected, the normal hierarchy bound is well below those shown
in Figure \ref{fig:Contour90} due to cancellations induced by the
marginalization over unknown phases. However, care must be taken when
interpreting this result as one hierarchy may be preferred over another
by anarchy. A numerical scan over an ensemble of mass eigenvalues,
without imposing mass squared differences from oscillation data, reveals
a general preference for the intermediate state to lie closer to the
heavier than the lighter one. My results agree qualitatively with
a similar scan done in \citep{Haba:2000be}. Still, this slight effect
does not suggest that anarchy favors the inverted hierarchy, which
is defined in terms of mass squared differences. These results are
relatively independent of the supplied weighting function, provided
that it is sufficiently non-singular.

\section{Conclusions\label{sec:Conclusion}}

Within the framework of neutrino mass anarchy, the distribution of
parameters and matrix elements must be treated statistically. Unlike
a study of neutrino mixing angles and phases, which depend only on
the invariant Haar measure, an analysis of mass eigenvalues/matrix
elements depends critically on arbitrary integration measures and
boundary conditions. For well behaved measures, the value of any one
matrix element may vary between about $0.01$ eV and $0.4$ eV at
90\% confidence. This is well within the reach of future experiments
and will be tightened with better knowledge of the overall neutrino
mass scale. However, these bounds are expanded for sufficiently divergent
measures, which may lead to a vanishing matrix element. This case
also renders the remaining matrix elements undefined, and thus removes
almost all predictability from the theory. I conclude that arbitrarily
small matrix elements are allowed within the anarchy framework, but
this seems counter-intuitive, given the universal mass matrix structure
assumed in the original formulation of the theory. 

This is more puzzling when put in the framework of realistic neutrino
data. $m_{ee}=0$ implies the normal hierarchy and practically forces
a $\mu-\tau$ like symmetry among the remaining elements \citep{MyLNV,StructureNuMassMatrixCpViolation,ElementsOfNuMassMatrix_AllowedRangesTextureZeros}.
While still consistent with anarchy (the divergent weighting imposes
the texture zero and frees all other elements to vary almost unconstrained
to yield the correct mass spectra and parameter relationships), this
seems like a lot of accidental structure for a structureless matrix!
The resolution lies in the ultraviolet completion of the theory that
produces the integration measure. That is, the underlying theory yields
well defined mass parameters, but by our limited knowledge of its
complicated structure, this information is communicated to low energies
as an ensemble of possible choices that must be treated statistically.
The nature of possible model classes select the weighting function
for us. An anarchy guaranteed texture zero implies a mechanism selecting
those textures from the ensemble of ultraviolet completions. This
is inconsistent with the anarchy principle from construction, as it
would require a corresponding symmetry mechanism and/or parameter
fine tuning. Of course, without knowing the details of neutrino mass generation it is impossible to gauge or select uncomfortable levels of tuning or when flavor structures must arise.  Still, one would not expect the metric power law parameter to venture too close the the $-6$ singularity.  To get a handle on this, it is reasonable to assume a 10\% or greater deviation which implies $m_{ee}\geq5\times10^{-3}$ eV at 90\% confidence.
$\beta\beta0\nu$ bounds below this level should indicate nontrivial
structure in the lepton flavor sector, new light LNV degrees of freedom
\citep{MyLowESeeSaw,Andre:SeeSawEnergyScaleLSNDAnomaly,Boris:CptCpCPhasesEffectsMajoranaProcesses,Wolfenstein:CpPropertiesMajoranaNuBB0n}
or the Dirac nature of the neutrino.

\begin{acknowledgments}
This work was motivated by questions raised at the PHENO 2008 symposium
on the smallest $\beta\beta0\nu$ rates allowed without imposed structure.
I thank Andre de Gouvea for useful insight into this topic and comments
on the original manuscript.  This paper was edited by Tina Jenkins.  This work is sponsored in part by the
US Department of Energy Contract DE-FG02-91ER40684.
\end{acknowledgments}
\bibliographystyle{apsrev}
\bibliography{/home/james/Work/MyReferences}

\end{document}